\documentclass[pre,showpacs,showkeys,preprint]{revtex4}
\begin{document}

\title{Interaction of $N$ solitons in the massive Thirring model and 
optical gap system: the complex Toda chain model}

\author{V. S. Shchesnovich}
\altaffiliation{On leave from The Division for 
Optical Problems in Information Technologies, National Academy of 
Sciences of Belarus, Zhodinskaya St. 1/2, 220141 Minsk, Belarus}
\email{valery@optoinform.bas-net.by}

\affiliation{Department of Mathematics and Applied Mathematics, 
University of Cape Town, Private Bag 7701 Rondebosch, South 
Africa}

\date{\today}

\begin{abstract}
Using the Karpman-Solov'ev quasiparticle approach for soliton-soliton 
interaction I show that the train propagation of $N$ well separated 
solitons of the massive Thirring model is described by the complex 
Toda chain with $N$ nodes. For the optical gap system a generalised  
(non-integrable) complex Toda chain is derived for description of the train 
propagation  of well separated gap solitons. These results are in favor 
of the recently proposed conjecture of universality of the complex Toda 
chain. 
\end{abstract}
\pacs{05.45.Yv, 42.65.Tg, 42.81.Dp}
\keywords{soliton-soliton interaction, soliton propagation, gap solitons in 
optical fibers, complex Toda chain} 
\maketitle

\section{Introduction}
Recently the complex Toda chain attracted much attention as a 
possible candidate for description of the pulse interactions
in integrable and non-integrable nonlinear evolution equations 
\cite{1,arn1,2,3,4,5,6,arn2,7,8}. For instance, it was shown that the 
complex Toda chain describes the soliton train propagation for all the 
nonlinear evolution equations associated with the NLS hierarchy \cite{7}. 
Quite recently the complex Toda chain was derived for the modified NLS 
equation~\cite{8}, an integrable generalisation of the NLS equation, which 
is  associated with the quadratic bundle. 

The complex Toda chain  is an integrable generalisation of the 
well-known real Toda chain (see, for instance, Refs.~\cite{2,3}). In 
Refs.~\cite{2,3,6} and \cite{8} the comparison of the complex Toda chain 
predictions with  the numerical solutions of the NLS and MNLS equations has 
been performed and a good agreement has been established for various choices 
of the initial parameters of the solitons in the train. 

It is noted that the complex Toda chain arises as an approximation of the 
evolution equations describing the inter-pulse interaction in the train 
comprised of well separated solitons with nearly equal amplitudes 
and velocities. The exponent of the (negative) separation between the 
solitons serves as the  small parameter for the asymptotic expansion and 
derivation of the complex Toda chain can be based either on the variational 
approach (see Ref.~\cite{arn2}) or on the adiabatic 
perturbation theory for solitons (see, for instance, Refs.~\cite{3,8}).  
However, as noted in Ref.~\cite{mik} the variational approach should be used 
with care. The approach based on the adiabatic perturbation theory is 
equivalent to the Karpman-Solov'ev quasiparticle method for the two-soliton 
interactions \cite{KS}. This approach was developed in 
Refs.~\cite{2,3,5,7,8}. 

If the nonlinear PDE is not integrable but possesses stable soliton 
solutions, then the train propagation of solitons is described by a 
generalised (non-integrable) complex Toda chain as it is pointed out in 
Ref.~\cite{arn2}.  
 
The complex Toda chain allows a rich class of asymptotic regimes of the 
soliton train propagation \cite{2,5,6}: \textit{i})~asymptotically free 
propagation of solitons, \textit{ii}) $N$-soliton bound states with the 
possibility of a quasi-equidistant propagation, \textit{iii})  mixed 
asymptotic regimes when part of the solitons form bound state(s) and the 
rest separate from them, \textit{iv}) regimes corresponding to the 
degenerate and singular solutions of the complex Toda chain. The rich 
variety of dynamical regimes of the  complex Toda chain indicates that it is 
a good candidate for analytical study of the soliton trains. Here I should 
point out that only few simple regimes are exhibited by the real Toda chain
\cite{RTC1,RTC2}, thus it is essential to have the complex Toda chain in 
description of the soliton trains. Moreover, the phase space of the complex 
Toda chain with $N$ nodes is $4N$-dimensional,  which is precisely the 
number of the real parameters in the train of $N$ solitons.
 
In the present paper I consider the $N$-soliton train propagation governed 
by two intimately related nonlinear PDEs, one of which is integrable and 
the other is not: the massive Thirring model of the classical field 
theory~\cite{thi,kuz} and the optical gap system 
\cite{ster1,crist,10,roma,stee,feng,brod2,scal,brod3,cheng,13,11,Rev, 
justify}. 
 
For the massive Thirring model I show that the train propagation of well 
separated solitons with nearly equal amplitudes and rapidities is governed 
by the complex Toda chain. Moreover, I derive a non-integrable 
generalisation of the complex Toda chain, which describes the train 
propagation of well separated gap solitons with nearly equal amplitudes and 
velocities. 
 
The gap soliton propagation through a grated optical fiber was 
manifested in recent experiments~\cite{eggl,brod1,tave1,tave2}. The results 
of~\cite{tave2} are of particular interest: there the \textit{multiple 
gap soliton formation} was observed. 

Localised solutions in nonlinear media with periodic band gaps have a great 
potential for technological applications.  One of the most important 
band gap structures in optics is given by an optical fiber with periodic 
index grating along  the axis. From the Floquet-Bloch theory of 
wave propagation in periodic structures it is known that there are 
forbidden frequency bands or band gaps for linear waves. On the other hand, 
nonlinear wave propagation in such structures is possible for the 
central frequency of the wave packet lying in the band gap. Such 
nonlinear wave is usually called gap soliton. 

The optical gap system was derived within the coupled mode 
approach for nonlinear wave propagation in optical fibers with 
grating (see, for instance, Ref.~\cite{11}). It reads
\begin{eqnarray} 
i\left({E_1}_T-{E_1}_X\right)+E_2+\left(|E_2|^2+\rho|E_1|^2\right)E_1&=&0, 
\nonumber\\ & &\label{OGS}\\ 
i\left({E_2}_T+{E_2}_X\right)+E_1+\left(|E_1|^2+\rho|E_2|^2\right)E_2&=&0, 
\nonumber \end{eqnarray} where $E_1$ and~$E_2$ are the slowly varying 
envelopes of two counter propagating waves coupled through the Bragg 
scattering induced by the grating (the linear cross-coupling terms), the 
nonlinear terms account for the self- and cross-phase modulation effects. 
The parameter~$\rho$ ($\rho>0$) at the self-phase modulation term may 
range up to infinity \cite{12}, in which case the optical gap system models 
dynamics in the nonlinear dual-core asymmetric coupler~\cite{13}. 
Setting  $\rho=0$ in the system (\ref{OGS}) one obtains the massive 
Thirring model. 

The gap solitons in optical fibers were theoretically studied in many 
works \cite{ster1,crist,10,roma,stee,feng,brod2,scal,brod3,cheng,13,11} (see 
also the latest review Ref.~\cite{Rev}). Recently, the relevance of the 
system (\ref{OGS}) for description of the optical gap solitons  was 
analytically and numerically validated~\cite{justify}. 
The general family of the   gap solitons was derived in~\cite{10} using the 
similarity with the massive Thirring model.  Recently it was shown that the 
gap soliton becomes unstable  when its amplitude grows above some fixed 
value \cite{barpeli,ross}.  

The soliton-like solutions, which are similar to 
the gap solitons, were found in nonlinear  diatomic 
lattices~\cite{kivs1,kivs2,ster3} and in the quadratic ($\chi^{(2)}$) 
materials  with a spatially periodic linear susceptibility 
(grating)~\cite{bury,kivs3,kivs4,cont1,pesc,he,cont2,cont3}. 
Also, it was  shown~\cite{ster3,kivs4} that, at some limit, the equations 
governing nonlinear wave propagation  in quadratic media with grating and in 
diatomic lattices are similar to the system (\ref{OGS}) though  the 
underlying physics is different.  

In the next section, section 2, I state the main results of the paper on the 
soliton train propagation for the massive Thirring model and optical gap 
system. The details of the derivation  are placed in the following sections: 
section 3 for the massive Thirring model and section 4 for the optical gap 
system. The last section contains discussion of the results and suggestions 
for further work. 

\section{Main results} 
Before formulate the main results I would like to remind some facts about 
the models under study.  Let us begin with the massive Thirring model of 
the classical field theory \cite{thi,kuz}: \begin{eqnarray}
i\left(v_t-v_x\right)+u+|u|^2v&=&0, \nonumber\\ 
& &\label{1}\\ 
i\left(u_t+u_x\right)+v+|v|^2u&=&0, \nonumber 
\end{eqnarray} 
where $u$ and $v$ are complex variables, $t$ and $x$ are the time and space 
co-ordinates, respectively.  The system (\ref{1})  is Lorentz-invariant:
\[
x\rightarrow \frac{x-\text{tanh}(y)t}{(1-\text{tanh}^2y)^{1/2}},\quad
t\rightarrow \frac{t-\text{tanh}(y)x}{(1-\text{tanh}^2y)^{1/2}},
\]
with $u$ and $v$ transforming as components of the Lorentz spinor,
\[
u \to e^{-y/2} u, \quad v \to e^{y/2}v.
\]
In the relativistic kinematics the parameter $y$  is called ``rapidity". 
(Rapidities of two consecutive Lorentz transformations simply add 
together.) 

The massive Thirring model is integrable by the inverse scattering transform 
method \cite{kuz}. For instance, its one-soliton solution can be written as 

\begin{eqnarray} 
v&=&-\frac{i\sin(2{\vartheta})\exp(-y/2+i\Theta)}{\cosh(z-i{\vartheta})},
\nonumber\\
& & \label{MTMsol}\\
u&=&\frac{i\sin(2{\vartheta})\exp(y/2+i\Theta)}{\cosh(z+i{\vartheta})}, 
\nonumber\end{eqnarray} where $0<{\vartheta}<\pi/2$ and 
\begin{equation} 
z=\sin(2{\vartheta})\text{cosh}(y)(x_o(t)-x),\quad 
\Theta=-\cot(2{\vartheta})\text{tanh}(y)z + \delta(t). 
\label{ztheta}\end{equation} 
The soliton has four independent real parameters: ${\vartheta}$, $y$, 
$x_o$, and $\delta$. The first two give the soliton amplitude and rapidity, 
while the rest two are the soliton position and central phase (the phase at 
$x=x_o$), respectively.  The position and phase parameters depend on time:
\begin{equation} 
\frac{\text{d}x_o}{\text{d}t}=\text{tanh}{y}, \quad 
\frac{\text{d}\delta}{\text{d}t}=-\cos(2{\vartheta})\text{sech}y. 
\label{xodelta}\end{equation}
The first equation defines the soliton velocity: $V = \text{ 
tanh}{y}$.

By a suitable Lorentz transformation the rapidity of a Thirring soliton can 
be put equal to zero and the soliton solution (\ref{MTMsol}) reduces to the 
quiescent soliton of the massive Thirring model. 

The following \textit{ansatz} is called the soliton train 
\begin{eqnarray}
v&=&\sum_{\alpha=1}^{N}-\frac{i\sin(2{\vartheta}_\alpha)\exp\left(-y_\alpha/2
+i\Theta_\alpha\right)}{\cosh(z_\alpha-i{\vartheta}_\alpha)},
\nonumber\\
& &\label{MTMtr}\\
u&=&\sum_{\alpha=1}^{N}\frac{i\sin(2{\vartheta}_\alpha)\exp\left(y_\alpha/2
+i\Theta_\alpha\right)}{\cosh(z_\alpha+i{\vartheta}_\alpha)},
\nonumber\end{eqnarray} 
where $z_\alpha$ and $\Theta_\alpha$ are given by formulae 
similar to (\ref{ztheta}) and (\ref{xodelta}).  It should be 
stressed that, for each soliton, all four soliton parameters in  formula 
(\ref{MTMtr}) are considered to be $t$-dependent. 
\medskip

\noindent{\bf The CTC for the MTM soliton train.}
Assume that the $N$-soliton train given by 
(\ref{MTMtr}) consists of well separated pulses with nearly equal amplitudes 
${\vartheta}_\alpha$ and rapidities $y_\alpha$, numerated by 
$\alpha=1,...,N$ in such a way that $x_{\alpha+1}-x_\alpha>0$ (here and 
below $x_\alpha$ denotes the position parameter ``$x_o$"  for the 
$\alpha$-th soliton). Mathematically these conditions are expressed as \[
|{\vartheta}_\alpha-\bar{{\vartheta}}|\ll \bar{\vartheta},\quad 
|y_\alpha-\bar{y}|\ll1,\quad |x_\alpha-x_{\alpha\pm1}|\gg 1,
\]
\begin{equation} 
|\sin(2{\vartheta}_\alpha)\cosh y_\alpha-\sin(2\bar{\vartheta})\cosh\bar{y}|
|x_\alpha-x_{\alpha\pm1}|\ll1.
\label{condits}\end{equation}
Here (and throughout the paper) $\bar{\vartheta}$ and $\bar{y}$ denote the 
averages:
\begin{equation}
\bar{{\vartheta}}=\frac{1}{N}\sum_{\alpha=1}^{N}{\vartheta}_\alpha,\quad
\bar{y}=\frac{1}{N}\sum_{\alpha=1}^{N}y_\alpha.
\label{aver}\end{equation}
Define the following new variables: a modified time 
\begin{equation}
\tau=\sin(2\bar{{\vartheta}})\text{sech}(\bar{y})t,
\label{modt}\end{equation} 
an average phase 
\begin{equation}
\bar{\delta}=-\cos(2\bar{{\vartheta}})\text{sech}(\bar{y})t,
\label{avdelta}\end{equation}
and the following complex variable for each soliton 
\begin{widetext}
\begin{equation} 
q_\alpha=-\sin(2\bar{{\vartheta}})\text{cosh}(\bar{y})x_\alpha
-i[\delta_\alpha-\bar{\delta}-\cos(2\bar{\vartheta}) 
\text{sinh}(\bar{y})x_\alpha +\alpha\pi] 
+ 2\alpha\ln[2\sin(2\bar{{\vartheta}})]. 
\label{qalpha}\end{equation}
Then  in the first order of the soliton overlap  parameter 
$\epsilon$, 
\begin{equation} \epsilon\simeq 
\exp\{-|\sin(2{\vartheta}_\alpha)\cosh(y_\alpha)x_\alpha 
-\sin(2{\vartheta}_{\alpha\pm1})\cosh(y_{\alpha\pm1})x_{\alpha\pm1}|\}, 
\label{paramet}\end{equation}
\end{widetext}
the following two statements are claimed:
\begin{enumerate}
\item the average values $\bar{\vartheta}$ and $\bar{y}$
do not depend on $t$;
\item evolution of the quantities $q_\alpha$, $\alpha=1,...,N$, is 
given by the complex Toda chain with $N$ nodes:
\begin{equation}
\frac{\text{d}^2q_\alpha}{\text{d}\tau^2}=e^{q_{\alpha+1}-q_\alpha}
-e^{q_\alpha-q_{\alpha-1}}, \quad \alpha=1,...,N,
\label{CTCforMTM}\end{equation}
where $\text{Re}\{q_{0}\}=\infty$ and $\text{Re}\{q_{N+1}\}=-\infty$ 
(i. e., $x_0=-\infty$ and $x_{N+1}=\infty$, see (\ref{qalpha})). 
\end{enumerate}
The set of inequalities (\ref{condits}) is similar to the 
inequalities for the NLS soliton train in Ref.~\cite{3}. 

Now I will formulate similar result for the train propagation of pulses 
governed by the optical gap system~\cite{11}: 
\begin{eqnarray} 
i\left({E_1}_t-{E_1}_x\right)+E_2+\left(|E_2|^2+\rho|E_1|^2\right)E_1&=&0, 
\nonumber\\ & &\label{ogs}\\ 
i\left({E_2}_t+{E_2}_x\right)+E_1+\left(|E_1|^2+\rho|E_2|^2\right)E_2&=&0, 
\nonumber \end{eqnarray}
The soliton solution of the optical gap system (\ref{ogs}) moving 
with the velocity $V=\tanh{y_o}$ reads \cite{10} 
\begin{eqnarray} 
\left(\begin{array}{c}E_1(x,t) \\ E_2(x,t)\end{array}\right)
=\frac{e^{i\psi(x,t)}}{[1+ \rho \cosh(2y_o)]^{1/2}} \left(\begin{array}{c} 
v(x,t)\\ u(x,t)\end{array}\right),
\label{gapsol}\end{eqnarray}
where ${v}$ and ${u}$ have the form of a Thirring soliton,  i.e., 
given by formulae (\ref{MTMsol})-(\ref{xodelta}) (with $y\to y_o$); the 
additional (nonlinear) phase $\psi$ is
\begin{equation} 
{\psi}=-\frac{2\rho\sinh(2y_o)} { 1+ \rho \cosh(2y_o) }
\text{arctan}(\tan{\vartheta}\tanh{z}), 
\label{onesol}\end{equation}
with $z$ as in equation (\ref{ztheta}).

Here it should be pointed out that the gap soliton becomes unstable when 
the soliton amplitude ${\vartheta}$ grows above certain threshold 
(${\vartheta}_{thr}\approx\pi/4$, see Ref.~\cite{barpeli}). This scenario is 
also possible for the train of gap solitons.  This instability is the result 
of the soliton-radiation interaction and is beyond the scope of the 
adiabatic approach. However, being interested in \textit{stable}  gap 
solitons, one can impose the condition  
${\vartheta}_\alpha<{\vartheta}_{thr}$, for all $\alpha=1,...,N$. 

The \textit{ansatz} I use for the train of \textit{well separated} gap 
solitons is given by application of the transformation (\ref{gapsol}) to the 
train of well separated Thirring solitons (in this case $y_o=\bar{y}$). 
Due to the inequalities (\ref{condits}), the additional phase $\psi_\alpha$  
of each soliton in the train can be approximated by formula (\ref{onesol}) 
with ${\vartheta}=\bar{\vartheta}$ and 
\mbox{$z_\alpha=\sin(2\bar{\vartheta})\text{cosh}(\bar{y})(x_\alpha - x)$.}

\medskip

\noindent{\bf The generalised CTC for the train of gap solitons.}
Assume that the train of $N$ gap solitons consists of well separated
pulses with nearly equal amplitudes ${\vartheta}_\alpha$ and 
rapidities $y_\alpha$ numerated by $\alpha=1,...,N$ in such 
a way that $x_{\alpha+1}-x_\alpha>0$ and that the conditions (\ref{condits}) 
are satisfied.  Associate the following variables with each gap soliton
\begin{widetext}
\[
Q_\alpha =-\sin(2\bar{{\vartheta}})\text{cosh}(\bar{y})x_\alpha
-i\{\delta_\alpha-\bar{\delta}
-[\cos(2\bar{\vartheta})-\mu\sin(2\bar{\vartheta})(y_\alpha-\bar{y})]
\text{sinh}(\bar{y})x_\alpha +\alpha\pi\}
\]
\begin{equation}
+2\alpha\ln[2\sin(2\bar{\vartheta})],
\label{qforOGS}\end{equation}
\end{widetext}
where
\[
\mu=\frac{4\rho\tanh(2\bar{y})}{\rho+\text{sech}
(2\bar{y})}\bar{{\vartheta}}. 
\] 
Define the modified time $\tau$ and average phase 
$\bar{\delta}$ as in formulae (\ref{modt}) and (\ref{avdelta}). Then, in the 
first order of the soliton overlap  parameter $\epsilon$ 
(\ref{paramet}), the following is claimed: 
\begin{enumerate}
\item the average values $\bar{\vartheta}$ and $\bar{y}$
do not depend on $t$;
\item evolution of the quantities $Q_\alpha$, 
$\alpha=1,...,N$, is given by the following generalised complex Toda chain 
with $N$ nodes: 
\begin{equation}
\frac{\text{d}^2Q_\alpha}{\text{d}\tau^2}
=(1+A_\rho)\Bigl(e^{Q_{\alpha+1}-Q_\alpha} 
- e^{Q_\alpha-Q_{\alpha-1}}\Bigr)
+B_\rho\Bigl(e^{Q^*_{\alpha+1}-Q^*_\alpha}
-e^{Q^*_\alpha-Q^*_{\alpha-1}}\Bigr),
\label{CTCforOGS}\end{equation}
where $\text{Re}\{Q_{0}\}=\infty$ and $\text{Re}\{Q_{N+1}\}=-\infty$. 

Equation (\ref{CTCforOGS}) is valid for \textit{arbitrary} values of the 
self-phase modulation parameter $\rho$. 
\end{enumerate}

Here $A_\rho$ and $B_\rho$ are $\rho$-dependent coefficients: 
\begin{eqnarray}
A_\rho &=& \frac{1}{2}\{\nu-\kappa\mu+i[\kappa(1+\nu)+\mu]\},
 \nonumber\\
& & \label{AB}\\
B_\rho&=& \frac{1}{2}\{\nu+\kappa\mu-i[\kappa(1+\nu)-\mu]\},
\nonumber\end{eqnarray}
with
\[
\kappa = \frac{\rho \tanh(2\bar{y})}{\rho +\text{sech}(2\bar{y})}
\frac{4\bar{\vartheta} - \sin(4{\vartheta})}{\sin^2(2{\vartheta})},\quad
\nu = \frac{4\rho(2\bar{{\vartheta}}\cot(2\bar{{\vartheta}})-1)}
{\rho+\text{sech}(2\bar{y})}. 
\]

Setting $\rho=0$ in (\ref{qforOGS}) and (\ref{CTCforOGS})  one obtains the 
complex Toda chain  for the soliton train of the massive Thirring model. 
 
\section{The CTC for the MTM soliton train}

I will use the adiabatic perturbation theory for derivation of the complex 
Toda chain. Recently the perturbation theory based on the Riemann-Hilbert 
problem was developed for the solitons of the massive Thirring 
model~\cite{9}. For instance, we have derived the following evolution 
equations for the soliton parameters in the adiabatic approximation: 
\begin{widetext}
\begin{equation} 
\frac{\text{d}{\vartheta}}{\text{d}t}=-\frac{1}{2\cosh{y}} 
\int\limits_{-\infty}^\infty \frac{\text{d}z}{\cosh(2z)+\cos(2{\vartheta})} 
\text{Re}\left\{e^z[r_\perp(k_1,z) +{r}^*_\perp(k^*_1,-z)]\right\}, 
\label{4a}\end{equation} 
\begin{equation} 
\frac{\text{d}y}{\text{d}t}=\frac{1}{\cosh{y}} \int\limits_{-\infty}^\infty 
\frac{\text{d}z}{\cosh(2z)+\cos(2{\vartheta})} 
\text{Im}\left\{e^z[r_\perp(k_1,z) +{r}^*_\perp(k^*_1,-z)] 
-2r_\parallel(z)\right\}, \label{4b}\end{equation}
\begin{equation}
\frac{\text{d}x_o}{\text{d}t}=\tanh{y}-
\frac{\text{Re}\{J\}}{2\sin^2(2{\vartheta})\cosh^2{y}},
\label{4c}\end{equation}
\begin{equation}
\frac{\text{d}\delta}{\text{d}t}=-\cos(2{\vartheta})\text{sech}y
-\frac{\text{Im}\{J\}+\tanh{y}\cot(2{\vartheta})
\text{Re}\{J\}}{2\sin(2{\vartheta})\cosh{y}}.
\label{4d}\end{equation}
Here
\[
J=\int\limits_{-\infty}^\infty\frac{\text{d}z}{\cosh(2z)+\cos(2{\vartheta})}
\Bigl(2\{2z[\cos(2{\vartheta})+i\sin(2{\vartheta})\tanh{y}]
+\sinh(2z)\}r_\parallel(z)
\]
\begin{equation}
+\{e^z[\cos(2{\vartheta})(1-2z)-2iz\sin(2{\vartheta})\tanh{y}]
+e^{-z}\}[r_\perp(k_1,z)-{r}^*_\perp(k^*_1,-z)]\Bigr).
\label{J}\end{equation}
\end{widetext}
The functions $r_\perp$ and~$r_\parallel$
are given by 
\begin{equation}
r_\perp(k_1,z,t)=\frac{ie^{-i\Theta}}{2}
\left(k_1\frac{\delta v}{\delta t}-k_1^{-1}\frac{\delta u}{\delta t}\right),
\label{5a}\end{equation}
\begin{equation}
r_\parallel(z,t)=\frac{i}{4}\left(\frac{\delta |v|^2}{\delta t}
-\frac{\delta |u|^2}{\delta t}\right),
\label{5b}
\end{equation}
where $k_1=-\exp\{-y/2-i{\vartheta}\}$. The ``variational"
derivatives denote fictitious evolution of the $u$ and $v$ as if 
under the action of the perturbation only (in other words, either $i\delta 
u/\delta t$  or $i\delta v/\delta t$  is nothing but the perturbation added 
to the  respective r.h.s. of the massive Thirring model (\ref{1})). 
Equations (\ref{4a})-(\ref{5b}) can also be derived via the perturbation 
theory for the Thirring solitons developed in Ref.~\cite{kauplakoba}.

From the point of view of the inverse scattering transform method
the \textit{ansatz} (\ref{MTMtr}) contains not only the $N$-soliton 
solution but also the contribution of radiation as well. However, due to 
large separations between the solitons in the train, the radiation component 
is negligible. Thus, I can use the adiabatic perturbation theory for 
derivation of evolution equations for the soliton parameters 
${\vartheta}_\alpha$, $y_\alpha$, $x_\alpha$, and $\delta_\alpha$. For the 
same reason, it is sufficient to consider the interaction between the 
neighboring pulses only (detailed discussion can be found in Ref.~\cite{3}). 

First, one should compute the perturbation functions defined in (\ref{5a}) 
and (\ref{5b}). Substitution of the \textit{ansatz} (\ref{MTMtr}) into 
the massive Thirring model (\ref{1}) and expansion of the cubic terms
leads to the following formulae for the perturbation-induced evolutions (I 
consider the interaction between the neighboring pulses only) 
\begin{eqnarray} 
i\frac{\delta v_\alpha}{\delta 
t}&=&-\sum_{\beta=\alpha\mp1}\left(|u_\alpha|^2v_\beta
+2\text{Re}\{u_\alpha{u}^*_\beta\}v_\alpha 
\right), \nonumber\\
& & \label{9}\\
i\frac{\delta u_\alpha}{\delta 
t}&=&-\sum_{\beta=\alpha\mp1}\left(|v_\alpha|^2u_\beta
+2\text{Re}\{v_\alpha{v}^*_\beta\}u_\alpha \right)
\nonumber\end{eqnarray} 
and, consequently, 
\begin{eqnarray} 
i\frac{\delta |v_\alpha|^2}{\delta 
t}&=&\sum_{\beta=\alpha\mp1} 2i|u_\alpha|^2
\text{Im}\{v_\alpha{v}^*_\beta\}, 
\nonumber\\
& &\label{10}\\
i\frac{\delta |u_\alpha|^2}{\delta 
t}&=&\sum_{\beta=\alpha\mp1} 2i|v_\alpha|^2
\text{Im}\{u_\alpha{u}^*_\beta\}. 
\nonumber\end{eqnarray} 

Before giving the perturbation-induced evolution of the soliton parameters 
some remarks must be maid on the details of the approximation due to the 
inequalities (\ref{condits}). The r.h.s.'s in (\ref{9})  
contain the small parameter $\epsilon$ (\ref{paramet}). I consider the 
soliton-soliton interaction in the first order with respect to $\epsilon$. 
Hence, due to the presence of the small parameter, the differences between 
the $\alpha$-th soliton amplitude and rapidity  and the average values of 
these quantities (given by (\ref{aver})) are negligible in the terms 
accounting for the inter-soliton interaction.  

Substitution of (\ref{9}) and (\ref{10}) into (\ref{5a}) and 
(\ref{5b}) and the result into (\ref{4a}) and (\ref{4b})  gives the 
following  equations for the amplitudes and rapidities 
\begin{equation} 
\frac{\text{d}{\vartheta}_\alpha}{\text{d}t}=\sum_{\beta=\alpha\mp1} 
\frac{2\sin^3(2\bar{\vartheta})}{\cosh\bar{y}}
e^{-|\Delta_{\alpha\beta}|}
\sin\Psi_{\alpha\beta}, 
\label{11a}\end{equation} 
\begin{equation}
\frac{\text{d}y_\alpha}{\text{d}t}=\sum_{\beta=\alpha\mp1}
\frac{4\text{sgn}(\Delta_{\alpha\beta})\sin^3(2\bar{{\vartheta}})}
{\cosh\bar{y}} e^{-|\Delta_{\alpha\beta}|}
\cos\Psi_{\alpha\beta},
\label{11b}\end{equation}
where 
\[
\Delta_{\alpha\beta}=\sin(2\bar{{\vartheta}})\cosh(\bar{y})(x_\beta-x_\alpha),
\]
\[
\Psi_{\alpha\beta}=\delta_\alpha-\delta_\beta-\cos(2\bar{{\vartheta}})
\sinh(\bar{y})(x_\alpha-x_\beta).
\]
Here $\exp(-|\Delta_{\alpha\beta}|)={\cal O}(\epsilon)$. It can 
be easily verified that equations (\ref{11a}) and (\ref{11b}) do not affect 
the average amplitude $\bar{{\vartheta}}$ and rapidity $\bar{y}$.

The evolution equations for $x_\alpha$ and $\delta_\alpha$ (see (\ref{4c})
and (\ref{4d})) are comprised of two  addends, which account for the  
unperturbed and perturbation-induced evolution (the latter contain the 
soliton overlap parameter $\epsilon$). Hence, one can neglect the terms 
accounting for the perturbation-induced evolution of these parameters as 
compared to their unperturbed evolution: 
\begin{equation} 
\frac{\text{d}x_\alpha}{\text{d}t}=\tanh{y_\alpha},\quad 
\frac{\text{d}\delta_\alpha}{\text{d}t}=-\cos(2{\vartheta}_\alpha)
\text{sech}{y_\alpha}. 
\label{12}\end{equation} 

Now everything is ready for derivation of the complex Toda chain. 
Let us differentiate the following quantity 
$-\Delta_{\alpha\beta}+i\Psi_{\alpha\beta}$ ($\beta =\alpha\pm1$): 
\[
\frac{\text{d}}{\text{d}t}(-\Delta_{\alpha\beta}+i\Psi_{\alpha\beta})
=\sin(2\bar{{\vartheta}})\text{sech}(\bar{y})[(y_\alpha+ 
2i{\vartheta}_\alpha) - (y_\beta+ 2i{\vartheta}_\beta)],
\]
where the second order terms in $({\vartheta}_\alpha-\bar{{\vartheta}})$ and 
$(y_\alpha-\bar{y})$ are dropped. On the other hand, from equations 
(\ref{11a}) and (\ref{11b}) one derives 
\begin{widetext}
\[ 
\frac{\text{d}}{\text{d}t}(y_\alpha+2i{\vartheta}_\alpha) 
=\sum_{\beta=\alpha\mp1}
\frac{4\text{sgn}(\Delta_{\alpha\beta})\sin^3(2\bar{{\vartheta}})}{ 
\cosh\bar{y}} \exp\{\text{sgn}(\Delta_{\alpha\beta}) 
(-\Delta_{\alpha\beta}+i\Psi_{\alpha\beta})\}, 
\] 
or using the numeration $x_{\alpha+1}-x_\alpha>0$, i.e. 
$\text{sgn}(\Delta_{\alpha,\alpha+1})>0$, for the solitons in the train: 
\begin{equation}
\frac{\text{d}}{\text{d}t}(y_\alpha+2i{\vartheta}_\alpha)=
\frac{4\sin^3(2\bar{{\vartheta}})}{\cosh\bar{y}}\left( 
\exp\left\{-\Delta_{\alpha\alpha+1} + 
i\Psi_{\alpha\alpha+1}\right\}-\exp\left\{\Delta_{\alpha\alpha-1} - 
i\Psi_{\alpha\alpha-1}\right\}\right). 
\label{15}\end{equation} 
\end{widetext}
Introduce an average phase 
\[
\bar{\delta}=-\cos(2\bar{{\vartheta}})\text{sech}(\bar{y})t,
\] 
and the following complex variables associated with each soliton 
\[
q_\alpha=-\sin(2\bar{{\vartheta}})\text{cosh}(\bar{y})x_\alpha
-i[\delta_\alpha-\bar{\delta}+\alpha\pi -\cos(2\bar{\vartheta}) 
\sinh(\bar{y})x_\alpha] + 2\alpha\ln[2\sin(2\bar{{\vartheta}})]. 
\]
(The average phase in the formula for $q_\alpha$ eliminates the 
constant phase gradient.) Then 
\begin{equation}
e^{\pm (q_{\alpha\pm1}-q_{\alpha})} = 
-4\sin^2(2\bar{\vartheta})\exp(\mp\Delta_{\alpha\alpha\pm1}\pm 
i\Psi_{\alpha\alpha\pm1}).
\label{17}\end{equation}
Differentiating $q_\alpha$ and neglecting the second order terms  I get
\begin{widetext}
\[
\frac{\text{d}q_\alpha}{\text{d}t}=
-\sin(2\bar{\vartheta})\cosh\bar{y}\tanh{y_\alpha}
-i[\sin(2\bar{\vartheta})\text{sech}(\bar{y})(2{\vartheta}_\alpha
-2\bar{{\vartheta}})+\cos(2\bar{\vartheta})
\text{sech}\bar{y}\tanh(\bar{y})(y_\alpha-\bar{y}) \]
\[
-\cos(2\bar{\vartheta})\sinh\bar{y}\tanh{y_\alpha}]
\]
\end{widetext}
(here the average phase $\bar\delta$ plays an important role for the 
expansion over ${\vartheta}_\alpha-\bar{\vartheta}$).
The second differentiation (with removal of the second order terms) gives: 
\begin{equation}
\frac{\text{d}^2q_\alpha}{\text{d}t^2}
=-\sin(2\bar{\vartheta})\text{sech}(\bar{y})\left(\frac{\text{d}
y_\alpha}{\text{d}t} 
+2i\frac{\text{d}{\vartheta}_\alpha}{\text{d}t}\right). 
\label{secondder}\end{equation} At the same time, as it follows from 
(\ref{15}) and (\ref{17}), \begin{equation} 
\frac{\text{d}}{\text{d}t}
(-y_\alpha-i(2{\vartheta}_\alpha-2\bar{{\vartheta}}))= 
\sin(2\bar{\vartheta})\text{sech}(\bar{y})
\left(e^{q_{\alpha+1}-q_\alpha}-e^{q_\alpha-q_{\alpha-1}}\right). 
\label{deriv1}\end{equation} 
Therefore, what is left is to introduce a new 
time variable 
\[ 
\tau = \sin(2\bar{\vartheta})\text{sech}(\bar{y})t. 
\]
Then  equations (\ref{secondder}) and (\ref{deriv1}) give the complex Toda 
chain for the train of Thirring solitons: 
\[
\frac{\text{d}^2q_\alpha}{\text{d}\tau^2}=e^{q_{\alpha+1}-q_\alpha}
-e^{q_\alpha-q_{\alpha-1}}, \quad \alpha=1,...,N.
\]
Here it is assumed that $\text{Re}\{q_{0}\}=\infty$ and 
$\text{Re}\{q_{N+1}\}=-\infty$ (i.e., $x_0=-\infty$ and $x_{N+1}=\infty$).

\section{The generalised CTC for the train of gap solitons}

In derivation of the complex Toda chain for the optical gap 
system I will use a one-to-one mapping, found recently in 
Ref.~\cite{9}, between the optical gap system (\ref{OGS}) and the following 
generalisation of the  massive Thirring model, the $\gamma$-system for 
short, 
\begin{eqnarray} i\left({\cal V}_t-{\cal V}_x\right)+{\cal U}+|{\cal 
U}|^2{\cal V}+\gamma_-\left(|{\cal V}|^2-|{\cal U}|^2\right){\cal V} &=&0, 
\nonumber\\ & &\label{gammasys}\\ 
i\left({\cal U}_t+{\cal U}_x\right)+{\cal 
V}+|{\cal V}|^2{\cal U}+\gamma_+\left(|{\cal U}|^2-|{\cal V}|^2\right){\cal 
U} &=&0. \nonumber 
\end{eqnarray} 
The transformation relating the two systems is as follows.  Let ${\cal 
U}(x,t)$ and ${\cal V}(x,t)$ be a solution of the $\gamma$-system 
(\ref{gammasys})  with the following  $ \gamma_{\pm}$: 
\begin{equation} 
\gamma_{\pm}= \frac{\rho e^{\pm 2y_o}}{1+ \rho \cosh(2y_o)},\label{gammas} 
\end{equation} 
then 
\begin{eqnarray} 
\left( \begin{array}{c}E_1(X,T) \\
E_2(X,T)\end{array} \right)
=\frac{e^{i\psi(x,t)}}{[1+ \rho \cosh(2y_o)]^{1/2}}
\left( \begin{array}{c} e^{-y_o/2}{\cal V}(x,t) \\ e^{y_o/2}{\cal U}(x,t)
\end{array} \right),
\label{map}\end{eqnarray}
where 
\begin{equation}
x=\frac{X-\text{tanh}(y_o)T}{[1-\text{tanh}^2(y_o)]^{1/2}}, 
\quad t=\frac{T-\text{tanh}(y_o)X}{[1-\text{tanh}^2(y_o)]^{1/2}},
\label{xt}
\end{equation}
is a solution to the optical gap system (\ref{OGS}) with the phase 
${\psi}$ given by the following system of equations (in the 
light-cone variables $\eta=(t+x)/2$ and $\xi=(t-x)/2$) 
\begin{equation}
\frac{\partial {\psi}}{\partial \eta}= \frac{1}{2}(\gamma_+-\gamma_-)|{\cal 
V}|^2 , \quad \frac{\partial {\psi}}{\partial \xi}=-\frac{1}{2}(\gamma_+
-\gamma_-)|{\cal U}|^2. 
\label{poten}
\end{equation}
(Note that the conservation of the number of particles, i. e., 
\[
\frac{\partial}{\partial \xi} |{\cal V}|^2 + 
\frac{\partial}{\partial \eta} |{\cal U}|^2 =0,
\]
ensures the compatibility of the equations for the phase $\psi$.)

Note that the co-ordinates are related via a Lorentz transformation.
The mapping (\ref{map}) can be verified by direct substitution into 
(\ref{OGS}) via simple calculations with the use of (\ref{gammas}), 
(\ref{xt}), and (\ref{poten}). 

The presented mapping is valid for \textit{arbitrary} solutions of the 
optical gap system. However, the importance of the mapping 
(\ref{gammas})-(\ref{poten}) stems from the fact that by choosing the 
\textit{quiescent}  Thirring soliton, 
\[ 
{\cal V}=-\frac{i\sin(2{\vartheta})e^{i\delta}}{\cosh(z-i{\vartheta})},\quad 
{\cal U}=\frac{i\sin(2{\vartheta})e^{i\delta}}{\cosh(z+i{\vartheta})}, \]
\[
z=-\sin(2{\vartheta})(x-x_o), \quad \frac{\text{d}\delta}{\text{d}t} 
= -\cos(2{\vartheta}),
\]
which is a solution to the system (\ref{gammasys}) due to $|{\cal V}|=|{\cal 
U}|$,  one can recover the optical gap soliton \textit{moving with any 
given velocity} $V=\tanh y_o$. In this case  one obtains formula 
(\ref{onesol}) for the additional phase phase $\psi$. Although the optical 
gap system is not Lorentz-invariant, still it makes sense to call $y_o$ 
``rapidity" of the gap soliton due to the transformation 
(\ref{gammas})-(\ref{poten}).

A train of $N$ well separated gap solitons moving with arbitrary 
average rapidity $y_o$ can always be represented via the transformation 
(\ref{gammas})-(\ref{poten}) as a  train of $N$ well separated 
\textit{almost quiescent}  Thirring solitons. Indeed, application of 
the mapping  (\ref{gammas})-(\ref{poten})  to equation (\ref{MTMtr}) with 
${\cal V}=v$ and ${\cal U}=u$, under the conditions $|y_\alpha|\ll 1$ and 
$\bar{y}=0$, yields the train of $N$ well separated gap solitons with nearly 
equal amplitudes and rapidities, where the {\em average} rapidity is equal 
to the given {\em arbitrary }  value $y_o$. Moreover, if one neglects the 
terms of order ${\cal O}(\epsilon)$, then the  additional phase 
$\psi_\alpha$ of each gap soliton in the train is determined by equation 
(\ref{onesol}) with evident changes: ${\vartheta}\to\bar{\vartheta}$, $z\to 
z_\alpha$ and $x_o\to x_\alpha$.
\medskip

Convenience of the $\gamma$-system (\ref{gammasys}) for the 
analytical study of gap solitons is based on the two following facts. 
Firstly, the quiescent Thirring soliton satisfies $|{\cal U}|=|{\cal 
V}|$. Hence the last terms in the $\gamma$-system are small if the solution 
under study is close to the quiescent Thirring soliton, or, in terms of the 
optical gap  system, the solution is close to the gap soliton. Secondly, 
the parameters $\gamma_\pm$ are bounded for all values of $y_o$ and $\rho$ 
(including $\rho=\infty$). Therefore the use of the equivalent 
$\gamma$-system allows one to apply the  perturbation theory developed for 
the train of almost quiescent Thirring solitons to the train of gap solitons 
for \textit{arbitrary} values of the self-phase modulation parameter 
$\rho$. 

Hence, derivation of the complex Toda chain for the gap 
soliton train can be done in much the same  way as  the derivation of 
the complex Toda chain for the train of Thirring solitons (more precisely, 
almost quiescent Thirring solitons).  The only difference is that there are 
additional small perturbations  given by the terms with 
$\gamma_\pm$ in (\ref{gammasys}), the $\gamma$-terms for short.  Let us 
first calculate their contribution to the  evolution of the soliton 
parameters and then calculate evolution of $q_\alpha$,  defined in a 
similar was as in the previous section, with account of these terms as well. 

Below I will take into account that gap soliton train is transformed by 
the mapping (\ref{gammas})-(\ref{poten}) with $y_o=\bar{y}$ to the train of 
almost quiescent Thirring solitons (in the variables ${\cal V}$ and ${\cal 
U}$). For instance, the latter train has $\bar{y}=0$ and $|y_\alpha|\ll1$.
Consider the $\alpha$-th soliton in such train. From (\ref{gammasys}) one 
gets  expanding the cubic terms:
\begin{widetext}
\[ 
i\frac{\delta {\cal V}_\alpha}{\delta t} 
=-\sum_{\beta=\alpha\mp1}\left(|{\cal U}_\alpha|^2{\cal V}_\beta
+2\text{Re}\{{\cal U}_\alpha{{\cal U}}^*_\beta\}{\cal V}_\alpha 
\right)-\gamma_-\left(|{\cal V}_\alpha{}|^2- |{\cal U}_\alpha{}|^2\right) 
{\cal V}_\alpha 
\] 
\[
-\sum_{\beta=\alpha\mp1}\gamma_-\left(2|{\cal V}_\alpha{}|^2
{\cal V}_\beta{}+{\cal V}_\alpha{}^2{{\cal V}}^*_\beta{}
-|{\cal U}_\alpha{}|^2{\cal V}_\beta{}-2\text{Re}\{{\cal U}_\alpha{} 
{{\cal U}}^*_\beta{}\}{\cal V}_\alpha{}\right). 
\] 
Let us separate the r.h.s. into two parts. The first part is just the same 
as in the case of the massive Thirring solitons, while the second, given by 
the following formula 
\[
i\left(\frac{\delta {\cal V}_\alpha}{\delta t}\right)_\gamma= 
-\gamma_-\left(|{\cal V}_\alpha{}|^2- |{\cal U}_\alpha{}|^2\right) 
{\cal V}_\alpha 
\]
\begin{equation}
-\sum_{\beta=\alpha\mp1}\gamma_-\left(2|{\cal V}_\alpha{}|^2v_\beta{} 
+{\cal V}_\alpha{}^2{{\cal V}}^*_\beta{}-|{\cal U}_\alpha{}|^2v_\beta{}
-2\text{Re}\{{\cal U}_\alpha{}
{{\cal U}}^*_\beta{}\}{\cal V}_\alpha{}\right),
\label{latter}\end{equation}
\end{widetext}
is due to the $\gamma$-perturbation and is specific for the gap soliton 
train only. I will detailly consider only the $\gamma$-perturbation since 
the first part is accounted just in the same way as for the train of 
almost quiescent Thirring solitons with the same resulting formulae. 

In formula (\ref{latter}) the first term on the r.h.s. is due to the 
self-interaction of the gap soliton (due to the $\gamma$-terms in 
(\ref{gammasys}), if the rapidity $y_\alpha\ne0$), the rest account for the 
inter-soliton interaction in the train. The latter terms  contain the small 
parameter $\epsilon$ defined in (\ref{paramet}). I will use the same 
approximation which has been used for the  derivation of the complex Toda 
chain for the massive Thirring model.  Additionally one  can neglect the 
difference between the modules of ${\cal V}_\alpha$ and ${\cal U}_\alpha$ 
when calculating the contribution from the inter-soliton interaction terms. 
This is because the inter-soliton interaction terms already contain the 
small parameter $\epsilon$ and $|y_\alpha|\ll1$, thus one can put 
$y_\alpha=0$ there. Then, the ``variational" derivatives simplify 
considerably: 
\begin{widetext}
\begin{equation} 
i\left(\frac{\delta {\cal V}_\alpha}{\delta 
t}\right)_\gamma =-\gamma_-\left(|{\cal V}_\alpha{}|^2- |{\cal 
U}_\alpha{}|^2\right) {\cal V}_\alpha 
-\sum_{\beta=\alpha\mp1}4\gamma_- \sin(\Psi_{\alpha\beta})\text{Im}
\{{\cal V}_{o\alpha}{\cal U}_{o\beta}\}{\cal V}_\alpha,
\label{26a}\end{equation}
\begin{equation}
i\left(\frac{\delta {\cal U}_\alpha}{\delta t}\right)_\gamma
=-\gamma_+\left(|{\cal U}_\alpha{}|^2- |{\cal V}_\alpha{}|^2\right)
{\cal V}_\alpha -\sum_{\beta=\alpha\mp1}4\gamma_+
\sin(\Psi_{\alpha\beta})\text{Im}
\{{\cal U}_{o\alpha}{\cal V}_{o\beta}\}{\cal U}_\alpha,
\label{26b}\end{equation}
\end{widetext}
where ${\cal V}_{o\alpha}=\exp\{-i\delta_\alpha\}{\cal V}_\alpha$,   
${\cal U}_{o\alpha}=\exp\{-i\delta_\alpha\}{\cal U}_\alpha$, and  
$\Psi_{\alpha\beta}=\delta_\alpha-\delta_\beta$.
Formula (\ref{26b}) 
obtains from (\ref{26a})  by the evident substitution ${\cal V}\rightarrow 
{\cal U}$,  ${\cal U}\rightarrow {\cal V}$ and 
$\gamma_-\rightarrow\gamma_+$. Note that from  formulae (\ref{26a})  and 
(\ref{26b})  it follows  that 
\[
\left(\frac{\delta |{\cal V}_\alpha|^2}{\delta t}\right)_\gamma=0,\quad 
\left(\frac{\delta |{\cal U}_\alpha|^2}{\delta t}\right)_\gamma=0.
\]

Consider first the contribution to evolution of the soliton parameters 
coming from the inter-soliton interaction $\gamma$-terms, i.e., 
the first terms on the r.h.s.'s in formulae (\ref{26a}) and (\ref{26b}). 
First, the perturbation functions given in (\ref{5a}) and (\ref{5b}) must be 
calculated. The inter-soliton interaction terms give the following 
contributions to the necessary functions: $r_\parallel(z_\alpha)=0$ and 
\begin{widetext}
\[ 
r_\perp(k_\alpha,z_\alpha)+{r}^*_\perp(k^*_\alpha,-z_\alpha)= 
-\sum_{\beta=\alpha\mp1}8i\frac{\sin^4(2\bar{{\vartheta}}) 
e^{-|\Delta_{\alpha\beta}|}\sin(\Psi_{\alpha\beta}) 
}{[\cosh(2z_\alpha)+\cos(2\bar{{\vartheta}})]^2}\sinh(2z_\alpha) 
\] 
\begin{equation}
\times\Bigl[(\gamma_+-\gamma_-)e^{z_\alpha}
+(\gamma_+e^{2i\bar{\vartheta}}-\gamma_-e^{-2i\bar{\vartheta}})e^{-z_\alpha}
\Bigr],
\label{ers}\end{equation}
\end{widetext}
where $k_\alpha=-\exp\{-y_\alpha-i{\vartheta}_\alpha\}$ and 
$\Delta_{\alpha\beta}=\sin(2\bar{{\vartheta}})(x_\beta-x_\alpha)$. 
(On the r.h.s. of (\ref{ers}) the terms of the second order in 
${\vartheta}_\alpha-\bar{{\vartheta}}$  are neglected due to the small 
multiplier $\exp(-|\Delta_{\alpha\beta}|)={\cal O}(\epsilon)$.) Now, 
substitution of the expression (\ref{ers}) into (\ref{4a}) and  (\ref{4b}) 
leads to the following contributions to evolution equations for 
${\vartheta}_\alpha$ and $y_\alpha$: 
\[
\left(\frac{\text{d}{\vartheta}_\alpha}{\text{d}t}\right)_\gamma =
0, 
\]
\[
\left(\frac{\text{d}y_\alpha}{\text{d}t}\right)_\gamma = 
-\sum_{\beta=\alpha\mp1} 4\kappa 
\sin^3(2\bar{{\vartheta}})e^{-|\Delta_{\alpha\beta}|} 
\sin\Psi_{\alpha\beta}, \]
where
\[
\kappa = \frac{\rho \tanh(2\bar{y})}{\rho +\text{sech}(2\bar{y})}
\frac{4\bar{\vartheta} - \sin(4{\vartheta})}{\sin^2(2{\vartheta})}.
\]
What concerns the other two parameters $x_\alpha$ and $\delta_\alpha$, 
similar as in section 3, the inter-soliton interaction is of 
order ${\cal O}(\epsilon)$ and its contribution to evolution of the soliton 
position and phase can be neglected as compared to their unperturbed 
evolution. 

Now let us consider the contribution to evolution of the soliton parameters 
coming from the self-interaction $\gamma$-terms, i.e.,  the first terms in 
(\ref{26a}) and (\ref{26b}). I get the following contributions to the 
functions in (\ref{5a}) and (\ref{5b}):
\begin{equation}
r_\parallel(z_\alpha)=0,\quad
r_\perp(k_\alpha,z_\alpha)+{r}^*_\perp(k^*_\alpha,-z_\alpha)=0,\quad
\label{self1}\end{equation}
\begin{widetext}
\begin{equation}
r_\perp(k_\alpha,z_\alpha)-{r}^*_\perp(k^*_\alpha,-z_\alpha)=
\frac{4i\sin^3(2{\vartheta}_\alpha)y_\alpha}
{[\cosh(2z_\alpha)+\cos(2{\vartheta}_\alpha)]^2}\Bigl[(\gamma_-
-\gamma_+)e^{z_\alpha}+(\gamma_-e^{-2i{\vartheta}_\alpha}
-\gamma_+e^{2i{\vartheta}_\alpha})e^{-z_\alpha}\Bigr]
\label{self2}\end{equation}
(here it is taken into account that terms of the second  order in $y_\alpha$ 
are negligible).  From (\ref{self1}) it follows that the contributions  from 
the self-interaction $\gamma$-terms to  evolution equations for the 
amplitude ${\vartheta}_\alpha$ and rapidity $y_\alpha$ vanish, while 
substitution of (\ref{self1}) and (\ref{self2})  into (\ref{J}) gives  
due to $|y_\alpha|\ll1$:
\[ 
J_\alpha=2\left\{(\gamma_- + 
\gamma_+)[2\sin^2(2{\vartheta}_\alpha)-2{\vartheta}_\alpha 
\sin(4{\vartheta}_\alpha)]+i(\gamma_--\gamma_+) 
2{\vartheta}_\alpha\sin^2(2{\vartheta}_\alpha)\right\} y_\alpha. 
\] 
\end{widetext}
Thus the contributions from the self-interaction 
$\gamma$-terms to evolution of $x_\alpha$ and $\delta_\alpha$ are determined 
by the following coefficients 
\[ 
\nu_\alpha=-\frac{\text{Re}\{J_\alpha\}}{2\sin^2(2{\vartheta}_\alpha)
y_\alpha}=(\gamma_-+\gamma_+)
[4{\vartheta}_\alpha\cot(2{\vartheta}_\alpha)-2], 
\] 
\[ 
\mu_\alpha=-\frac{\text{Im}\{J_\alpha\}}{2\sin^2(2{\vartheta}_\alpha)
y_\alpha}=(\gamma_+ -\gamma_-)2{\vartheta}_\alpha. 
\] 
I can take just the 
average values of these coefficients (denoted below as $\nu$ and $\mu$),  
because the following combinations $\nu_\alpha y_\alpha$ and $\mu_\alpha 
y_\alpha$ will enter the evolution equations for the soliton parameters and 
$|y_\alpha|\ll1$. In other words, one can throw away the terms of the second 
order in ${\vartheta}_\alpha-\bar{\vartheta}$ and $y_\alpha$. Taking into 
account the definition of the $\gamma_\pm$, where $y_o=\bar{y}$, I obtain: 
\[
\nu=\frac{4\rho(2\bar{{\vartheta}}\cot(2\bar{{\vartheta}})-1)}
{\rho+\text{sech}(2\bar{y})},\quad 
\mu=\frac{4\rho\tanh(2\bar{y})}{\rho+\text{sech}
(2\bar{y})}\bar{{\vartheta}}. 
\]

Let us collect all the contributions, i.e., the terms same as for the  
Thirring soliton train (see equations (\ref{11a})-(\ref{12})) and those 
accounting for the $\gamma$-terms, and write down the corresponding  
evolution equations for the parameters of the $\alpha$-th gap soliton. 
They read:
\begin{equation} 
\frac{\text{d}{\vartheta}_\alpha}{\text{d}t}=\sum_{\beta=\alpha\mp1} 
2\sin^3(2\bar{{\vartheta}})e^{-|\Delta_{\alpha\beta}|} 
\sin\Psi_{\alpha\beta}, 
\label{40}\end{equation} 
\begin{equation} 
\frac{\text{d}y_\alpha}{\text{d}t}=\sum_{\beta=\alpha\mp1} 
4\sin^3(2\bar{{\vartheta}}) e^{-|\Delta_{\alpha\beta}|}
[\text{sgn}(\Delta_{\alpha\beta}) \cos\Psi_{\alpha\beta} 
-\kappa\sin\Psi_{\alpha\beta}], 
\label{41}\end{equation} 
\begin{equation} 
\frac{\text{d}\delta_\alpha}{\text{d}t}
=-\cos(2{\vartheta}_\alpha)+\mu\sin(2{\vartheta}){y_\alpha},
\label{42}\end{equation}
\begin{equation}
\frac{\text{d}x_\alpha}{\text{d}t}=(1+\nu)y_\alpha.
\label{43}\end{equation}
Here 
\[
\Delta_{\alpha\beta}=\sin(2\bar{{\vartheta}})(x_\beta-x_\alpha),
\quad \Psi_{\alpha\beta}=\delta_\alpha-\delta_\beta.
\]
It is easy to see that the averages $\bar{\vartheta}$ and $\bar{y}$ are not 
affected by equations (\ref{40}) and (\ref{41}). Equations 
(\ref{40})-(\ref{43}) are similar to those for the (almost quiescent) 
Thirring solitons, however there are additional terms in the evolution 
equations for ${y}_\alpha$, $\delta_\alpha$, and $x_\alpha$. 

Let us now derive the generalised complex Toda chain corresponding to 
equations (\ref{40})-(\ref{43}). As the derivation is quite similar to that 
for the \textit{quiescent} Thirring solitons  I will skip some details. 
As in the case of the massive Thirring model, introduce the modified time 
\[
\tau=\sin(2\bar{{\vartheta}})t,
\] 
an average phase 
\[
\bar{\delta}=-\cos(2\bar{{\vartheta}})t,
\]
and the complex  variables $q_\alpha$ for each soliton: 
\begin{equation}
q_\alpha=-\sin(2\bar{{\vartheta}})x_\alpha
-i(\delta_\alpha-\bar{\delta}+\alpha\pi)+2\alpha
\ln[2\sin(2\bar{{\vartheta}})].
\label{qforgap}\end{equation}
Differentiating $q_\alpha$ and throwing away the second order terms
one obtains
\[
\frac{\text{d}q_\alpha}{\text{d}\tau}=-\{(1+\nu+i\mu)y_\alpha 
+i[2{\vartheta}_\alpha-2\bar{{\vartheta}}]\}. 
\]
Differentiation of this formula gives
\begin{widetext}
\[
\frac{\text{d}^2q_\alpha}{\text{d}\tau^2}
=-\sum_{\beta=\alpha\mp1}4\text{sgn}
(\Delta_{\alpha\beta})\sin^2(2\bar{{\vartheta}}) 
\exp\left\{\text{sgn}(\Delta_{\alpha\beta}) 
(-\Delta_{\alpha\beta}+i\Psi_{\alpha\beta})\right\} \]
\[
-(\nu+i\mu)\text{Re}\sum_{\beta=\alpha\mp1}4\text{sgn}
(\Delta_{\alpha\beta})\sin^2(2\bar{{\vartheta}}) 
\exp\left\{\text{sgn}(\Delta_{\alpha\beta}) 
(-\Delta_{\alpha\beta}+i\Psi_{\alpha\beta})\right\} \]
\[
+\kappa(1+\nu+i\mu)\text{Im}\sum_{\beta=\alpha\mp1}4\text{sgn}
(\Delta_{\alpha\beta})\sin^2(2\bar{{\vartheta}}) 
\exp\left\{\text{sgn}(\Delta_{\alpha\beta}) 
(-\Delta_{\alpha\beta}+i\Psi_{\alpha\beta})\right\}. \]
\end{widetext}
Taking into account the numeration of the solitons in the train, which is 
given by $x_{\alpha+1}-x_\alpha>0$ or $\Delta_{\alpha\alpha+1}>0$, 
and the following identity
\[
4\sin^2(2\bar{{\vartheta}})\exp\{\pm(-\Delta_{\alpha\alpha\pm1}
+i\Psi_{\alpha\alpha\pm1})\}=-e^{\pm(q_{\alpha\pm1}-q_\alpha)},
\]
I obtain a generalised complex Toda chain for the train of $N$ well  
separated gap solitons with nearly equal amplitudes and rapidities:
\begin{equation}
\frac{\text{d}^2q_\alpha}{\text{d}\tau^2}
=(1+A_\rho)\Bigl(e^{q_{\alpha+1}-q_\alpha} - e^{q_\alpha-q_{\alpha-1}}\Bigr)
+B_\rho\Bigl(e^{q^*_{\alpha+1}-q^*_\alpha}-e^{q^*_\alpha-q^*_{\alpha-1}}\Bigr),
\label{pertCTC}\end{equation}
where $A_\rho$ and $B_\rho$ are $\rho$-dependent coefficients: 
\[
A_\rho = \frac{1}{2}\{\nu-\kappa\mu+i[\kappa(1+\nu)+\mu]\},
\]
\[
B_\rho = \frac{1}{2}\{\nu+\kappa\mu-i[\kappa(1+\nu)-\mu]\}.
\]
As usual, $\text{Re}\{q_{0}\}=\infty$ and $\text{Re}\{q_{N+1}\}=-\infty$. 

Though in equation (\ref{pertCTC}) and in the definition of  
$q_\alpha$ (\ref{qforgap}) I still have the variables $\tau$, $x_\alpha$, 
$\delta_\alpha$, and $\bar{\delta}$ defined through the co-ordinates  $x$ 
and $t$ (see formula (\ref{xt})), it is easy to reverse to the 
co-ordinates $X$ and $T$ of  the optical gap system (\ref{OGS}).  Indeed, to 
this end one should use the  transformation (\ref{xt}) (with $y_o=\bar{y}$) 
for the position $x_\alpha$ and the central phase $\delta_\alpha$ of the gap 
soliton (the phase at $X=X_\alpha$): 
\begin{equation} 
x_\alpha = \text{cosh}(\bar{y})[X_\alpha -\tanh(\bar{y})T],
\label{postr}\end{equation}
\begin{equation}  
\delta_\alpha = 
[-\cos(2{\vartheta}_\alpha)+\mu\sin(2\bar{\vartheta})y_\alpha]t 
=[-\cos(2{\vartheta}_\alpha) + \mu\sin(2\bar{\vartheta})y_\alpha]
\text{cosh}(\bar{y})[T - \tanh(\bar{y})X_\alpha].
\label{phasetr}\end{equation} 
Also one must use the time transformation 
$\text{d}T=\text{cosh}(\bar{y})\text{d}t$ in the definition of $\tau$ and 
the average phase $\bar{\delta}$: 
\[ 
\tau=\sin(2\bar{{\vartheta}})\text{sech}(\bar{y})T,\quad \bar\delta = 
-\cos(2\bar{\vartheta})\text{sech}(\bar{y})T. \]
Now it is evident that, if equation (\ref{postr}) 
is used in the definition of $q_\alpha$ (\ref{qforgap}), the term
linear in $T$  will not  give contribution neither to the 
difference $q_\alpha -q_\beta$ nor to the second derivative of $q_\alpha$, 
hence it can be dropped. Further, notice that from the r.h.s. of equation 
(\ref{phasetr}) only the term linear in $T$ will appear in $q_\alpha$  
(\ref{qforgap}), if one simply changes the time: $t\to 
\text{sech}(\bar{y})T$. Hence, the term proportional to $X_\alpha$ in 
(\ref{phasetr})  \textit{must be subtructed} from the central phase 
$\delta_\alpha$. In doing so, one can neglect the difference between 
${\vartheta}_\alpha$ and $\bar{\vartheta}$ due to the inequalities 
(\ref{condits}) and that evolution of ${\vartheta}_\alpha$ is of order 
${\cal O}(\epsilon)$ (i.e., we throw away the  second order terms from the 
second derivative of $q_\alpha$). Thus we have arrived precisely at the 
quantity $Q_\alpha$ given by equation (\ref{qforOGS}), where the shift of 
the soliton rapidities is taken into account: $y_\alpha\to\bar{y}+y_\alpha$. 
 Therefore the result of section 2 is proven. 
 
\section{Comments}

The complex Toda chain model proves to be an universal model for 
the adiabatic description of the train interaction/propagation of solitons 
in nonlinear PDEs. Indeed, it was shown to describe the train propagation of 
pulses in the nonlinear PDEs of the whole NLS hierarchy \cite{7} (i.e., the 
PDEs associated with the familiar Zakharov-Shabat spectral problem 
\cite{ZS,bookZS}). More recently, the complex Toda chain was derived for the 
soliton train of the modified NLS equation \cite{8}. This PDE is associated 
with the quadratic bundle, also known as the Wadati-Konno-Ichikawa spectral 
problem \cite{WKI}. 

In this paper, the complex Toda chain is shown to describe the soliton 
train propagation in the massive Thirring model. Note that, as it is 
mentioned in Ref.~\cite{8}, the massive Thirring model is just another 
representative of the modified NLS hierarchy. Thus the complex Toda chain 
arises in the adiabatic description of the soliton trains in the hierarchy 
of nonlinear integrable PDEs associated with the quadratic bundle as well. 
This is in favor of the universality of the complex Toda chain. 

In construction of the perturbation theory for the massive Thirring model
we have used the associated Riemann-Hilbert problem \cite{9}. The usage of 
the Riemann-Hilbert problem allows one to develop the perturbation theory in 
an unified way for the entire hierarchy (see, for instance, Ref.~\cite{JMP}, 
where this was done for the vector NLS hierarchy). Moreover, the 
perturbation-induced evolution equations for the spectral data  have one and 
the same form for \textit{all} integrable PDEs (one can compare the results 
of Refs.~\cite{RH1,RH2}). This gives  a possibility to prove the 
universality of the complex Toda chain using the approach based on the 
Riemann-Hilbert problem. This is one of the directions for future work. 

In view of recent experimental observation \cite{tave2} of the multiple 
gap soliton formation in optical fibers with index grating, it is 
important to have an analytical approach for decription of interaction of 
optical gap solitons. In this paper an analytical approach is developed for 
the train interaction/propagation of gap solitons: the adiabatic propagation 
of $N$ gap solitons with nearly equal amplitudes and velocities is governed 
by a generalised complex Toda chain with $N$ nodes. 

Here I should point out that,  due to non-integrability of the optical gap 
system, the train of gap solitons  may become unstable.  This instability is 
the result of the soliton-radiation interaction and is beyond the adiabatic 
approximation. However, the gap soliton is stable if the soliton amplitude 
lies below the instability threshold (see for details Ref.~\cite{barpeli}). 
In that case, the generalised complex Toda chain (\ref{CTCforOGS}) can be 
applied. Though in accordance with the Ref.~\cite{arn2} the generalised 
complex Toda chain  is not integrable, it is just a finite dimensional 
dynamical system and can be investigated by the standard techniques.  
Moreover, in accordance with discussion of Ref.~\cite{7}, one can 
systematically include various additional perturbations of the optical gap 
system into the complex Toda chain.  This is the direction for  further 
work. 

\begin{acknowledgements}
The author gratefully acknowledges many stimulating discussions 
with Professors E.V. Doktorov and V.S. Gerdjikov.  Also, the author is 
indebted to Professor V.S. Gerdjikov for his critical reading of the 
manuscript. Some part of this work was done during the RCP 264 Conference 
(June 2000, Montpellier, France) and the author wishes to thank  the 
organisers, Professors J.-G. Caputo and P. Sabatier, for their support. 
This work has been supported by  the National Research Foundation of South 
Africa. 
\end{acknowledgements}

\end{document}